\documentclass[10pt,journal]{IEEEtran}
\usepackage{amsmath,amssymb,bm,color,url}
\usepackage[noadjust]{cite}
\ifCLASSINFOpdf
  \usepackage[pdftex]{graphicx}
   \DeclareGraphicsExtensions{.pdf,.jpeg,.png}
\else
  \usepackage[dvips]{graphicx}
  \DeclareGraphicsExtensions{.eps}
\fi

\newcommand{\Tr}{\mathrm{Tr}}

\DeclareMathOperator{\rank}{rank}
\DeclareMathOperator{\diag}{diag}

\begin{document}

% ------
\title{On the Detection of Passive Eavesdroppers in the MIMO Wiretap Channel}
\author{Amitav~Mukherjee, \IEEEmembership{Member, IEEE,} and A.~Lee~Swindlehurst, \IEEEmembership{Fellow, IEEE}
\thanks{A. Mukherjee is with the Wireless Systems Research Lab of Hitachi America, Ltd., Santa Clara, CA 95050-2639, USA (e-mail: \tt{amitav.mukherjee@hal.hitachi.com}).}
\thanks{A. L. Swindlehurst is with the Center for Pervasive Communications and Computing,
University of California, Irvine, CA 92697-2625, USA  (e-mail: \tt{swindle@uci.edu}).}% <-this % stops a space
\thanks{This work was presented in part at IEEE ICASSP 2012, and was supported by the National Science Foundation under grant CCF-0916073.}
}
\maketitle
%\pagestyle{empty}
%\thispagestyle{empty}
%\IEEEpeerreviewmaketitle
\begin{abstract}
The classic MIMO wiretap channel comprises a passive eavesdropper that attempts to intercept communications between an authorized transmitter-receiver pair, each node being equipped with multiple antennas. In a dynamic network, it is imperative that the presence of an eavesdropper be determined before the transmitter can deploy robust secrecy-encoding schemes as a countermeasure. This is a difficult task in general, since by definition the eavesdropper is passive and never transmits. In this work we adopt a method that allows the legitimate nodes to detect the passive eavesdropper from the local oscillator power that is inadvertently leaked from its RF front end. We examine the performance of non-coherent energy detection and optimal coherent detection, followed by composite GLRT detection methods that account for unknown parameters. Numerical experiments demonstrate that the proposed detectors allow the legitimate nodes to increase the secrecy rate of the MIMO wiretap channel.
\end{abstract}


\section{Introduction}
The broadcast characteristic of the wireless propagation medium makes it difficult to shield transmitted signals from unintended recipients. This is especially true in multiple-input multiple-output (MIMO) systems with multi-antenna nodes, where the increase in communication rate to the legitimate receiver is offset by the enhanced interception capability of eavesdroppers. A three-terminal network consisting of a legitimate transmitter-receiver pair and a passive eavesdropper where each node is equipped with multiple antennas is commonly referred to as the MIMO {\em wiretap} or MIMOME channel. The extent of information leakage in such systems is captured by the notion of {\em secrecy capacity} at the physical layer, which
quantifies the maximal rate at which a transmitter can reliably send a secret
message to the receiver, with the eavesdropper being completely unable to decode
it. Maximizing the achievable secrecy rate at the physical layer can therefore complement encryption performed at higher layers \cite{Mukherjee_Surv}. The secrecy capacity of the MIMO wiretap channel has been studied in
\cite{GoelN08}-\cite{Hassibi11}, for example.

In the burgeoning literature on the MIMO wiretap channel, a number of transmit precoding techniques have been proposed to improve the channel secrecy rate by exploiting knowledge of either the instantaneous realizations or statistics of the channel to the eavesdropper \cite{GoelN08}-\cite{Khisti10}. However, the question of how the legitimate transmitter acquires a passive eavesdropper's CSI has yet to be answered satisfactorily. The authors have previously proposed precoding schemes for the MIMO wiretap channel when the eavesdropper's CSI is completely unknown in \cite{Mukherjee09,Mukherjee11}. More importantly, it is imperative that the presence of a passive eavesdropper be determined \emph{before} the transmitter can deploy robust secrecy-encoding schemes as a countermeasure. This is a difficult task in a dynamic wireless network, since by definition the eavesdropper is passive and never transmits. To our best knowledge, the problem of determining the potential presence of passive eavesdroppers in the wiretap channel has not been addressed previously.

In this work we propose a scheme that allows the legitimate nodes to detect the passive eavesdropper from the local oscillator power that is inadvertently leaked from its RF front-end even when in reception mode. This technique was recently proposed in \cite{Wild05,Milstein06,Milstein10} for spectrum sensing in single-antenna cognitive radios (CR) to avoid interfering with primary receivers under AWGN channels. We generalize this technique to MIMO channels in a wiretap scenario for a variety of detectors based on energy detection, matched filtering, and composite tests. We then investigate how the proposed detection algorithms allow the legitimate nodes to increase the MIMO secrecy rate of the channel. The eavesdropper detection problem is essentially analogous to very low-SNR multi-antenna spectrum sensing in cognitive radio networks. While the majority of prior work on multi-antenna CR spectrum sensing aim to detect a single-antenna primary transmitter in zero-mean white Gaussian noise, in this work we explicitly consider the detection of a full-rank signal of interest in non-zero-mean Gaussian noise.

The remainder of this work is organized as follows. The MIMO wiretap channel with a passive eavesdropper is introduced in Sec.~\ref{sec:SysModel}. Coherent and noncoherent methods of detecting Eve's presence are characterized in Sec.~\ref{sec:EveDet}. Composite tests accounting for unknown noise and leakage parameters are examined in Sec.~\ref{sec:GLRTs}. The optimization of Eve's parameters given knowledge of the tests to detect her presence are briefly discussed in Sec.~\ref{sec:EveStrats}, followed by numerical results and conclusions in Sec.~\ref{sec:sim} and Sec.~\ref{sec:Conclu}, respectively.

\emph{Notation}:
We will use $\mathcal{CN}(\mathbf{c},\mathbf{Z})$ to denote a circularly symmetric complex
Gaussian distribution with mean $\mathbf{c}$ and covariance matrix $\mathbf{Z}$, and $\mathcal{N}(\mathbf{c},\mathbf{Z})$ for the real-valued counterpart.  Furthermore, we let
$\mathbb{E}\{\cdot\}$ denote expectation, $(\cdot)^T$ is the transpose, $(\cdot)^H$ is the Hermitian
transpose, $\Re$ represents the real part, $\Tr\{\cdot\}$ is the
trace operator, $\rank(\cdot)$ is the matrix rank, $\diag\{\mathbf{a}\}$ is a diagonal matrix with $\mathbf{a}$ on the main diagonal, $\lambda_i(\mathbf{A})$ is the $i^{th}$ ordered eigenvalue of $\mathbf{A}$ in descending order, ${}_2{F_1}\left( \cdot,\cdot;\cdot;\cdot \right)$ is the hypergeometric function, and $\mathbf{I}$ is an identity matrix of appropriate dimension.

\section{System Model}\label{sec:SysModel}
\subsection{Network Model}
We consider a multi-user network with an $N_a$-antenna transmitter (Alice), an $N_b$-antenna
receiver (Bob), and an unauthorized eavesdropper (Eve) with $N_e$ antennas. When Alice is transmitting to Bob and Eve is listening in the vicinity, the received signals at Bob and Eve at time instant $t$ are given by
\begin{eqnarray}
  {{\mathbf{y}}_b}\left(t \right) &=& \sqrt{d_{ab}^{-\alpha}}{{\mathbf{H}}_{ba}}\mathbf{x}\left(t \right) + {\mathbf{n}}_b\left(t \right) \hfill \\
  {{\mathbf{y}}_e}\left(t \right) &=& \sqrt{d_{ae}^{-\alpha}}{{\mathbf{H}}_{ea}}\mathbf{x}\left(t \right) + {\mathbf{n}}_e\left(t \right), \hfill
\end{eqnarray}
where $\mathbf{x}\left( t \right)\in {\mathbb{C}^{{N_a} \times 1}}$ is the confidential information signal, ${\mathbf{H}}_{ba}\in \mathbb{C}^{{N_b} \times N_a},{\mathbf{H}}_{ea}\in \mathbb{C}^{{N_e} \times N_a}$ are the deterministic and invariant complex MIMO channels from Alice, the distances from Alice to Bob and Eve are $d_{ab}>0$ and $d_{ae}>0$, respectively, and $\alpha$ is the path-loss exponent.
The additive complex Gaussian noise vectors are assumed to be independent, spatially uncorrelated, and distributed as ${\mathbf{n}}_b\left(t \right) \sim \mathcal{CN}\left( {{\mathbf{0}},{\sigma_b^2\mathbf{I}}} \right),{\mathbf{n}}_e\left(t \right) \sim \mathcal{CN}\left( {{\mathbf{0}},{\sigma_e^2\mathbf{I}}} \right)$. An average power constraint is imposed on Alice's transmit covariance matrix ${\mathbf{Q}} = \mathbb{E}\left\{ {{\mathbf{x}\left(t \right)}{{\mathbf{x}\left(t \right)}^H}} \right\}$ in the form of $\Tr \left( {\mathbf{Q}} \right) \leq {P_a}$.  Irrespective of the potential presence of Eve, both Alice and Bob are assumed to have perfect knowledge of the main channel ${\mathbf{H}}_{ba}$, which can be attained by the use of conventional training methods. If the input signal ${\mathbf{x}}$ is drawn from a Gaussian distribution, the instantaneous MIMO secrecy rate \cite{Khisti10} for fixed channels when Eve is present is given by
\begin{equation}\label{eq:MIMOSecRate}
%R_s=\left[ {I\left( {{\mathbf{x}} ;{\mathbf{y}}_b } \right) - I\left( {{\mathbf{x}} ;{\mathbf{y}}_e } \right)} \right]^{+}
\begin{split}
R_{s,i}&={\log _2}\left| {{\mathbf{I}} + d_{ab}^{-\alpha}\sigma_b^{-2}{{\mathbf{H}}_{ba}}{\mathbf{QH}}_{ba}^H} \right| \\
&\quad{-}\: {\log _2}\left| {{\mathbf{I}} + d_{ae}^{-\alpha}\sigma_e^{-2}{{\mathbf{H}}_{ea}}{\mathbf{QH}}_{ea}^H} \right|.
\end{split}
\end{equation}

\begin{figure}[htbp]
\centering
\includegraphics[width=\linewidth]{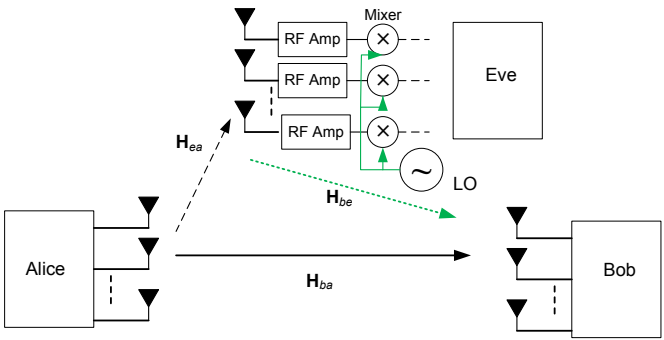}
\caption{MIMO wiretap channel with RF front end of Eve highlighted.}
\label{fig:Network}
\end{figure}

The wiretap channel is depicted in Fig.~\ref{fig:Network}. The fundamental procedure of detecting the passive node Eve is as follows.  We assume all three nodes possess either heterodyne or direct-conversion transceivers. A general impairment in such receivers is that a small portion of the local oscillator (LO) signal back-propagates to the antenna ports and leaks out, even when in passive reception mode \cite{Hamilton91}. While the LO leakage signal power is on the order of -50 to -90 dBm from a single antenna port, the LO leakage signal is boosted when multiple RF chains are present as in the MIMO wiretap setting, and is consequently easier to detect.

Therefore, we assume that Alice periodically ceases data transmission in order to allow \emph{both} herself and Bob to independently sense the radio environment, followed by a joint fusion of their individual decisions. Since the sensing algorithm and process is assumed to be identical at both Alice and Bob, to avoid repetition we focus on the local detection process at Bob in the sequel. The binary hypothesis test at Bob during these `silent' periods is
\begin{align}\label{eq:Bobhypo}%{*{20}{l}}
  {{H_0:}}&{{{\mathbf{y}}_b}\left( t \right) = \sqrt{d_{ab}^{-\alpha}}{{\mathbf{H}}_{ba}}{{\mathbf{w}}_l}\left( t \right)+{{\mathbf{n}}_b}\left( t \right)} \\
  {{H_1:}}&{{{\mathbf{y}}_b}\left( t \right) = \sqrt{d_{be}^{-\alpha}}{{\mathbf{H}}_{be}}{{\mathbf{s}}_l}\left( t \right) + \sqrt{d_{ab}^{-\alpha}}{{\mathbf{H}}_{ba}}{{\mathbf{w}}_l}\left( t \right)+{{\mathbf{n}}_b}\left( t \right)}\nonumber
\end{align}
where ${\mathbf{H}}_{be}\in \mathbb{C}^{{N_b} \times N_a}$ is the complex MIMO leakage channel from Eve to Bob who are separated by distance $d_{be}$. The LO leakage signals from Alice and Eve are represented by ${{\mathbf{w}}_l}\left( t \right)$ and ${{\mathbf{s}}_l}\left( t \right)$, respectively. Eq. \eqref{eq:Bobhypo} can also be used to model a distributed network of single-antenna sensors that report their observations over orthogonal channels to a fusion center in order to detect the presence of Eve.

The aggregate LO leakage signal from Eve is
\begin{equation}
{{\mathbf{s}}_l}\left( t \right) = {\left[ {\begin{array}{*{20}{c}}
  {{s_1}\left( t \right)}& \ldots &{{s_{{N_e}}}\left( t \right)}\end{array}} \right]^T}.
\end{equation}
We model the LO leakage signal from Eve's $i^{th}$ antenna port as an unmodulated frequency tone \cite{Milstein10}:
\begin{equation}
 {s_i}\left( t \right) = {A_i}\cos \left( {\omega t + {\theta _i(t)}} \right),
\end{equation}
where $A_i$ is the amplitude, $\omega$ is the LO frequency, and $\theta_i$ is an arbitrary time-varying phase. Similarly, the LO leakage signal from Alice is ${{\mathbf{w}}_l}\left( t \right) = {\left[ {\begin{array}{*{20}{c}}
{{w_1}\left( t \right)}& \ldots &{{w_{{N_a}}}\left( t \right)}\end{array}} \right]^T}$, where
\begin{equation}
{w_i}\left( t \right) = {B_i}\cos \left( {{\tilde\omega}t + {\xi _i(t)}} \right),
\end{equation}
where $B_i$ is the amplitude, ${\tilde\omega}$ is Alice's LO frequency, and $\xi_i$ is an arbitrary phase coefficient.

\subsection{Secrecy Rate Performance}
We consider the following signal transmission model. The overall data transmission period is split into blocks of $T$ channel uses. At the beginning of each block, Alice and Bob independently sense the radio environment for the presence of Eve. If the consensus is that Eve is absent, then for the remaining $T-1$ channel uses in that block Alice designs her input covariance $\mathbf{Q}$ to maximize the conventional MIMO rate to Bob via waterfilling \cite{ICASSP12}. If Eve is determined to be present, Alice acquires the statistics of her channel ${{\mathbf{H}}_{ea}}$ and optimizes $\mathbf{Q}$ by splitting her transmit resources between data and an artificial jamming signal such that the expected value of the MIMO secrecy rate for that block is maximized \cite{GoelN08}. The block duration $T$ is assumed to be long enough in order to invoke information-theoretic random coding arguments.

Define ${{P}_{dc}}$ and ${{P}_{fc}}$ as the overall consensus detection and false alarm probabilities derived via an arbitrary fusion rule from the local decisions at Alice and Bob.
For the commonly used AND and OR decision fusion rules, ${{P}_{dc}}$ can be defined as
\begin{equation}\label{eq:Decision_fusion}
{P_{dc}} = \left\{ {\begin{array}{*{20}{c}}
  {P_D^1P_D^2}&{\textrm{AND}{\text{ rule}}} \\
  { { {1 - \prod\nolimits_{i = 1}^2 {\left( {1 - P_D^i} \right)} }} }&{\textrm{OR}{\text{ rule}}}
\end{array}} \right.
\end{equation}
where $P_D^i,$ $\:i=1,2,$ are the local detection probabilities at Alice and Bob that are derived in subsequent sections.
If Eve is modeled as being present in a particular transmission block with a prior probability $\beta$ that is unknown to Alice/Bob, the expected value of the MIMO secrecy rate for an arbitrary block is written as
\begin{eqnarray}
  {{\bar R}_s} &=& {R_b}{{P}_{dc}}\left( {1 - \beta } \right) + {R_s}{{P}_{dc}}\beta  + \left( {{R_b} - {R_e}} \right)\left( {1 - {{P}_{dc}}} \right)\beta  \nonumber\\
   &&{+}\: {{\tilde R}_b}{{P}_{fc}}\left( {1 - \beta } \right),
\end{eqnarray}
where $R_s$ is the ergodic MIMO secrecy rate, $R_e$ is the information rate leaked to Eve upon missed detection, and ${\tilde R}_b$ is the sub-optimal rate to Bob when some resources are mistakenly allocated for secrecy encoding by Alice.

\subsection{Eavesdropper Detection}\label{sec:EveDet}
 The authors in \cite{Wild05} mainly focus on the use of a coherent matched filter detector \cite{KayVolII} for determining the presence of the primary receiver at a single-antenna cognitive radio. However, the matched filter approach requires phase synchronization at Bob as well as estimation of ${\mathbf{H}}_{be}$, which is exceedingly difficult given the very low LO leakage power. Park \emph{et al.} propose noncoherent envelope detection in the \emph{frequency} domain by applying a discrete Fourier transform (DFT) to the down-converted and sampled received signal \cite{Milstein06,Milstein10}, also in a single-antenna setting.
% The motivation behind this approach is the well-known fact that the DFT of a complex frequency tone concentrates the signal energy (and not the additive white noise) around a single frequency bin, whereas in the time domain both the signal and noise power are spread uniformly over all samples.
In this work we focus on multi-antenna detection in the time domain, and assume that Eve's LO frequency $\tilde\omega$ (or a good estimate of it) is known \emph{a priori} to both legitimate terminals during downconversion to baseband. If Eve employs a direct-conversion receiver, then as a worst-case scenario her LO frequency can be assumed to be known exactly since it is optimal for her to set $\tilde\omega = \omega$. The same is true if all terminals have an identical heterodyne architecture.%, therefore without loss of generality we assume $\tilde w = w$.%, for e.g., if the eavesdropper is in fact an idle user waiting for service in the system.

After downconverting and sampling, the hypothesis test at Bob based on $M$ discrete-time vector observations is
\begin{equation}
\begin{split}
  H_0:{{\mathbf{y}}_b}\left[ n \right] &= {{\mathbf{m}}_A}\left[ n \right] + {{\mathbf{n}}_b}\left[ n \right],\label{eq:H0}\\
  H_1:{{\mathbf{y}}_b}\left[ n \right] &= {{\mathbf{m}}_E}\left[ n \right] + {{\mathbf{m}}_A}\left[ n \right] + {{\mathbf{n}}_b}\left[ n \right],
\end{split}
\end{equation}
for $n = 0, \ldots ,M - 1,$ where
\begin{eqnarray*}
%\mathbf{m}_A\left[ n \right] &=& \sqrt{d_{ab}^{-\alpha}} {\mathbf{H}}_{ba}{{\mathbf{w}}_d}\left[ n \right]\\
\mathbf{m}_A\left[ n \right] &=& \sqrt{d_{ab}^{-\alpha}} {\mathbf{H}}_{ba}{{\mathbf{w}}_d}\left[ n \right]
\\ \mathbf{m}_E\left[ n \right] & = & \sqrt{d_{be}^{-\alpha}} \mathbf{H}_{be}{{\mathbf{s}}_d}\left[ n \right]\\
%\mathbf{m}_E\left[ n \right] &=& \sqrt{d_{be}^{-\alpha}} \mathbf{H}_{be}{{\mathbf{s}}_d}\left[ n \right] \\
\mathbf{w}_d\left[ n \right] &=& \left[ \begin{array}{*{10}{c}}
  {B_1} e^{\left( {j{\tilde{\omega}}n + {\xi _1[n]}} \right)}& \ldots & B_{N_a}e^{\left( {j{\tilde{\omega}} n + {\xi _{{N_a}}[n]}} \right)}
\end{array} \right]^T\\
\mathbf{s}_d\left[ n \right] &=& \left[ \begin{array}{*{10}{c}}
  {A_1}e^{\left( {j{\omega}n + {\theta _1[n]}} \right)}& \ldots & A_{N_e}e^{\left( {j{\omega}n + {\theta _{{N_a}}[n]}} \right)}
\end{array} \right]^T.
\end{eqnarray*}
The deterministic MIMO channels ${\mathbf{H}}_{ba}$ and ${\mathbf{H}}_{be}$ are assumed to be constant during the detection process. It is assumed that Bob's own leakage signal is removed and does not contaminate the detection process \cite{Milstein10}. The received signal has the following multivariate normal distributions:
\begin{equation}\label{eq:yb_multivnormal}
\begin{array}{*{20}{l}}
  {{{\mathbf{y}}_b}\left[ n \right] \sim \mathcal{CN}\left( {{{\mathbf{m}}_A}\left[ n \right],\sigma _b^2{\mathbf{I}}} \right)}&{{\text{under }}{H_0}} \\
  {{{\mathbf{y}}_b}\left[ n \right] \sim \mathcal{CN}\left( {{{\mathbf{m}}_E}\left[ n \right] + {{\mathbf{m}}_A}\left[ n \right],\sigma _b^2{\mathbf{I}}} \right)}&{{\text{under }}{H_1}}
\end{array}
\end{equation}

For convenience we aggregate the samples into a ($N_b \times M$) observation matrix
\begin{equation}{{\mathbf{Y}}_b} = \left[ {\begin{array}{*{20}{c}}
  {{{\mathbf{y}}_b}\left[ 0 \right]}& \ldots &{{{\mathbf{y}}_b}\left[ {M - 1} \right]}
\end{array}} \right]
\end{equation}
which follows a matrix-variate normal distribution \cite{James64} under both hypotheses:
\begin{equation}\label{eq:Yb_matrixnormal}
\begin{array}{*{20}{l}}
  {{{\mathbf{Y}}_b} \sim \mathcal{CN}\left( {{{\mathbf{M}}_A},\sigma _b^2{\mathbf{I}}} \right)}&{{\text{under }}{H_0}} \\
  {{{\mathbf{Y}}_b} \sim \mathcal{CN}\left( {{{\mathbf{M}}_E} + {{\mathbf{M}}_A},\sigma _b^2{\mathbf{I}}} \right)}&{{\text{under }}{H_1}}
\end{array}
\end{equation}
where we define
\begin{align}
{{\mathbf{M}}_A} &= \left[ {{{\mathbf{m}}_A}\left[ 0 \right]}, \ldots, {{{\mathbf{m}}_A}\left[ M-1 \right]}\right]\\
{{\mathbf{M}}_E} &= \left[ {{{\mathbf{m}}_E}\left[ 0 \right]}, \ldots, {{{\mathbf{m}}_E}\left[ M-1 \right]}\right] \label{eq:M_E}.
\end{align}

\section{Noncoherent and Coherent Detection}\label{sec:ED_MF}
\subsection{Energy Detection}\label{sec:ED}
Energy detection (ED) is a low-complexity noncoherent technique that obviates the need to estimate the leakage signal parameters and channels, and only requires an accurate estimate of the background noise variance $\sigma _b^2$ \cite{KayVolII}.
 %in (\ref{eq:H0})-(\ref{eq:H1}).
 The ED test statistic is given by
\begin{equation}
{T_{ED}}\left( {{{\mathbf{Y}}_b}} \right) = \operatorname{Tr} \left( {{\mathbf{Y}}_b^H{{\mathbf{Y}}_b}} \right) = \sum\limits_{n = 0}^{M - 1} {{{\left\| {{{\mathbf{y}}_b}\left[ n \right]} \right\|}^2}}.
\end{equation}
The ED hypothesis test compares the test statistic to a threshold $\eta$ to determine the presence of Eve:
\begin{equation}
 {T_{ED}}\left( {{{\mathbf{Y}}_b}} \right) \mathop \gtrless \limits_{{{H}_0}}^{{{H}_1}} \eta \; ,
\end{equation}
where $\eta$ is determined by a pre-specified probability of false alarm constraint $P_{FA}$.

From (\ref{eq:yb_multivnormal}), under both hypotheses ${T_{ED}}\left( {{{\mathbf{Y}}_b}} \right)$ has a noncentral chi-square distribution, since it is the sum of the squares of $2MN_b$ real and independent nonzero-mean Gaussian random variables:
\begin{equation}
\begin{array}{*{20}{c}}
  {{H_0}:}&{{T_{ED}}\left( {{{\mathbf{Y}}_b}} \right) \sim \frac{{\sigma_b^2}}{2}\chi_{2M{N_b}}^{'2}\left( {{\lambda _0}} \right)} \\
  {{H_1}:}&{{T_{ED}}\left( {{{\mathbf{Y}}_b}} \right) \sim \frac{{\sigma _b^2}}{2}\chi_{2M{N_b}}^{'2}\left( {{\lambda _1}} \right)}
\end{array}
\end{equation}
with associated noncentrality parameters
\begin{eqnarray*}
{\lambda _0} &=& \left( {{2 \mathord{\left/ {\vphantom {2 {\sigma _b^2}}} \right.
 \kern-\nulldelimiterspace} {\sigma _b^2}}} \right)\operatorname{Tr} \left( {\Re \left\{ {{\mathbf{M}}_A^T{{\mathbf{M}}_A}} \right\}} \right)\\
 {\lambda _1} &=&\left( {{2 \mathord{\left/
 {\vphantom {2 {\sigma _b^2}}} \right.
 \kern-\nulldelimiterspace} {\sigma _b^2}}} \right)\operatorname{Tr} \left( {\Re \left\{ {{{\left( {{{\mathbf{M}}_E} + {{\mathbf{M}}_A}} \right)}^T}\left( {{{\mathbf{M}}_E} + {{\mathbf{M}}_A}} \right)} \right\}} \right),
 \end{eqnarray*}
  respectively.
Under the null hypothesis, ${T_{ED}}\left( {{{\mathbf{Y}}_b}} \right)$ has the density function
 \begin{equation*}
 {f_T}\left( {t;{H_0}} \right) = \frac{{{e^{ - \left( {\frac{{{\lambda _0} + 2{t \mathord{\left/
 {\vphantom {t {\sigma _b^2}}} \right.
 \kern-\nulldelimiterspace} {\sigma _b^2}}}}{2}} \right)}}}}{{\sigma _b^2}}{\left( {\frac{{2t}}{{\sigma _b^2{\lambda _0}}}} \right)^{\frac{{M{N_b} - 1}}{2}}}{I_{M{N_b} - 1}}\left( {\sqrt {\frac{{2t{\lambda _0}}}{{\sigma _b^2}}} } \right)
 \end{equation*}
 and the probability of false alarm is calculated as
 \begin{equation}
 {P_{FA}} = {Q_{M{N_b}}}\left( {\sqrt {{\lambda _0}} ,\sqrt {\frac{{2\eta }}{{\sigma _b^2}}} } \right),
  \end{equation}
where ${Q_k}\left( {a,b} \right) = {a^{1 - k}}\int_b^\infty  {{t^k}{e^{ - \tfrac{{{t^2} + {a^2}}}{2}}}{I_{k - 1}}\left( {at} \right)} dt$ is the generalized Marcum $Q$-function, and $I_{k}$ is the modified Bessel function of the first kind of order $k$ \cite{KayVolII,MarcumInverse}. Similarly, the probability of detection is
 \begin{equation}
 {P_{D}} = {Q_{M{N_b}}}\left( {\sqrt {{\lambda _1}} ,\sqrt {\frac{{2\eta }}{{\sigma _b^2}}} } \right).
  \end{equation}
The value of the threshold $\eta$ that corresponds to a particular $P_{FA}$ can be computed by empirically evaluating the Marcum $Q$-function, or from the approximate inversion of the Marcum $Q$-function \cite{MarcumInverse}.

\subsection{Optimal Detector}\label{sec:OptDetector}
As an alternative to energy detection, we now consider the optimal Neyman-Pearson detector when all parameters of the leakage signals are assumed to be known to Bob. While un-realizable in practice, the optimal coherent detector provides an upper bound on the detection performance of any possible test. From (\ref{eq:yb_multivnormal})-(\ref{eq:Yb_matrixnormal}), the likelihood function under the null hypothesis is
\begin{equation}
\begin{gathered}
  f\left( {{{\mathbf{Y}}_b};{H_0}} \right) = \prod\limits_{n = 0}^{M - 1} {f\left( {{{\mathbf{y}}_b}\left[ n \right];{H_0}} \right)}  \hfill \\
   = \prod\limits_{n = 0}^{M - 1} {\frac{1}{{{{\left( {\pi \sigma _b^2} \right)}^{{N_b}}}}}\exp \left[ { - \frac{{{{\left( {{{\mathbf{y}}_b}\left[ n \right] - {{\mathbf{m}}_A}\left[ n \right]} \right)}^H}\left( {{{\mathbf{y}}_b}\left[ n \right] - {{\mathbf{m}}_A}\left[ n \right]} \right)}}{{\sigma _b^2}}} \right]}  \hfill \nonumber\\
   = \frac{1}{{{{\left( {\pi \sigma _b^2} \right)}^{M{N_b}}}}}\exp \left[ { - \frac{{\operatorname{Tr} \left\{ {{{\left( {{{\mathbf{Y}}_b} - {{\mathbf{M}}_A}} \right)}^H}\left( {{{\mathbf{Y}}_b} - {{\mathbf{M}}_A}} \right)} \right\}}}{{\sigma _b^2}}} \right] \hfill \\
\end{gathered}
\end{equation}
with the corresponding log-likelihood function
\begin{equation}\label{eq:L0}
\begin{split}
{\mathcal{L}_0}\left( {{{\mathbf{Y}}_b}} \right) &=  - M{N_b}\ln \left( {\pi \sigma _b^2} \right) \\
&\quad{-}\: \frac{1}{{\sigma _b^2}}\operatorname{Tr} \left\{ {{{\left( {{{\mathbf{Y}}_b} - {{\mathbf{M}}_A}} \right)}^H}\left( {{{\mathbf{Y}}_b} - {{\mathbf{M}}_A}} \right)} \right\}.
\end{split}
\end{equation}
Define ${{{\mathbf{M}}_1} \triangleq {{\mathbf{M}}_E} + {{\mathbf{M}}_A}}$. Under the alternative hypothesis $H_1$, a similar analysis yields
\begin{eqnarray*}
f\left( {{{\mathbf{Y}}_b};{H_1}} \right) &=& \frac{1}{{{{\left( {\pi \sigma _b^2} \right)}^{M{N_b}}}}} \nonumber\\
&&{\times}\exp \left[ { - \frac{{\operatorname{Tr} \left\{ {{{\left( {{{\mathbf{Y}}_b} - {{\mathbf{M}}_1} } \right)}^H}\left( {{{\mathbf{Y}}_b} - {{\mathbf{M}}_1}} \right)} \right\}}}{{\sigma _b^2}}} \right],
\end{eqnarray*}
\begin{eqnarray}\label{eq:L1}
{\mathcal{L}_1}\left( {{{\mathbf{Y}}_b}} \right) &=&  - M{N_b}\ln \left( {\pi \sigma _b^2} \right) \nonumber\\
&&{-}\: \frac{1}{{\sigma _b^2}}\operatorname{Tr} \left\{ {{{\left( {{{\mathbf{Y}}_b} - {{\mathbf{M}}_1} } \right)}^H}\left( {{{\mathbf{Y}}_b} - {{\mathbf{M}}_1} } \right)} \right\}.
\end{eqnarray}

The optimal Neyman-Pearson test compares the log-likelihood ratio to a threshold that corresponds to a particular $P_{FA}$:
\begin{equation}
{{\mathcal{L}_1}\left( {{{\mathbf{Y}}_b}} \right)}-{{{\mathcal{L}_0}\left( {{{\mathbf{Y}}_b}} \right)}} \mathop \gtrless \limits_{{{H}_0}}^{{{H}_1}}  \varepsilon'.
\end{equation}
Simple manipulations lead to the following test statistic:
\begin{equation}
{T_{op}}\left( {{{\mathbf{Y}}_b}} \right) = \operatorname{Tr} \left\{ {\Re \left( {{\mathbf{M}}_E^H{{\mathbf{Y}}_b}} \right)} \right\} \mathop \gtrless \limits_{{{H}_0}}^{{{H}_1}}  \varepsilon,
\end{equation}
where $\varepsilon  = \frac{1}{2} \sigma _b^2\varepsilon ' + 0.5\operatorname{Tr} \left\{ {{\mathbf{M}}_E^H\left( {{{\mathbf{M}}_E} + {{\mathbf{M}}_A}} \right) + {\mathbf{M}}_A^H{{\mathbf{M}}_E}} \right\}$. Therefore, the optimal detection rule is observed to be a replica-correlator or equivalently a matched filter, which is the expected outcome for detecting a known complex deterministic signal in Gaussian noise \cite{KayVolII}.

Next, we note that the test statistic is distributed as
\[\begin{array}{*{20}{c}}
  {{H_0}:}&{{T_{op}}\left( {{{\mathbf{Y}}_b}} \right)\sim\mathcal{N}\left( {\Re \left( {\Tr \left\{ {{\mathbf{M}}_E^H{{\mathbf{M}}_A}} \right\}} \right),\frac{{\sigma _b^2}}{2}\Tr \left\{ {{\mathbf{M}}_E^H{{\mathbf{M}}_E}} \right\}} \right)} \\
  {{H_1}:}&{{T_{op}}\left( {{{\mathbf{Y}}_b}} \right)\sim\mathcal{N}\left( {\Re \left( {\Tr \left\{ {{\mathbf{M}}_E^H{{\mathbf{M}}_1}} \right\}} \right),\frac{{\sigma _b^2}}{2}\Tr \left\{ {{\mathbf{M}}_E^H{{\mathbf{M}}_E}} \right\}} \right)}
\end{array}\]
from which we can derive the probabilities of detection and false alarm as
\begin{align}
{P_{FA}} &= Q\left( \frac{{\varepsilon  - \Re \left( {\Tr \left\{ {{\mathbf{M}}_E^H{{\mathbf{M}}_A}} \right\}} \right)}}{\sqrt{\frac{{\sigma _b^2}}{2}\Tr \left\{ {{\mathbf{M}}_E^H{{\mathbf{M}}_E}} \right\}}} \right)\label{eq:OptCohDetPFA}\\
{P_{D}} &= Q\left( \frac{{\varepsilon  - \Re \left( {\Tr \left\{ {{\mathbf{M}}_E^H{{\mathbf{M}}_1}} \right\}} \right)}}{\sqrt{\frac{{\sigma _b^2}}{2}\Tr \left\{ {{\mathbf{M}}_E^H{{\mathbf{M}}_E}} \right\}}} \right).
\end{align}
It is evident that the threshold value $\varepsilon$ that corresponds to a target false-alarm probability can be computed from \eqref{eq:OptCohDetPFA} as $\varepsilon = {\sqrt{\frac{{\sigma _b^2}}{2}\Tr \left\{ {{\mathbf{M}}_E^H{{\mathbf{M}}_E}} \right\}}} Q^{-1}({P_{FA}}) + \Re \left( {\Tr \left\{ {{\mathbf{M}}_E^H{{\mathbf{M}}_A}} \right\}} \right)$.

\section{Detection Under Unknown Parameters}\label{sec:GLRTs}
Thus far we have studied the energy detector, which does not require any information of the leakage parameters, and the optimal replica-correlator which assumes all parameters are known. To do better than ED, we can treat the leakage signal and channel parameters of Eve as unknown deterministic parameters to be estimated at Bob, and pose generalized likelihood ratio tests (GLRT) for these cases. The GLRT is a constant false-alarm rate detector which has featured prominently in the spectrum sensing literature \cite{Gazor10}-\cite{Soltanmohammadi13}, with various assumptions about the signal model. \cite{Gazor10}-\cite{Li10} consider the detection of rank-1 signals, \cite{Tugnait12} considers a test statistic based on the DFT of the received signal, and no performance analysis of the GLRT is given in \cite{Zeng10,Vilar11}. Alternatives to GLRTs with unknown parameters are the blind test in \cite{Zhao12} based on non-parametric empirical characteristic functions, and a heuristic test statistic based on the cross-correlation among signals at all antenna pairs \cite{Orooji11}. However, the test in \cite{Zhao12} cannot be characterized analytically, and \cite{Orooji11} assumes a particular channel autocorrelation model such as Clarke's or Jake's, which is not applicable when the signal of interest is a very low power sinewave as in our case.
In the sequel, we continue to assume that the leakage channel ${{{\mathbf{H}}_{ba}}}$ and related signal parameters from Alice are completely known at Bob \cite{ICASSP12}, and possibly to Eve as well.

\subsection{Unknown noise variance}\label{sec:GLRT1}
We begin with the case where both the effective leakage channels are changing slowly enough to have been determined in previous epochs, but the background noise variance in the current test epoch $\sigma _b^2$ is unknown, possibly due to time-varying interference. Following the standard derivation of the GLRT \cite{KayVolII}, we first compute the maximum likelihood estimates (MLEs) of $\sigma _b^2$ under the two competing hypotheses from the derivatives of (\ref{eq:L0})-(\ref{eq:L1}):
\begin{align}
  \hat \sigma _{b|{H_0}}^2 =& \frac{{\operatorname{Tr} \left\{ {{{\left( {{{\mathbf{Y}}_b} - {{\mathbf{M}}_A}} \right)}^H}\left( {{{\mathbf{Y}}_b} - {{\mathbf{M}}_A}} \right)} \right\}}}{{M{N_b}}} \label{eq:noisevarhat_H0} \\
  \hat \sigma _{b|{H_1}}^2 =& \frac{{\operatorname{Tr} \left\{ {{{\left( {{{\mathbf{Y}}_b} - {{\mathbf{M}}_1}} \right)}^H}\left( {{{\mathbf{Y}}_b} - {{\mathbf{M}}_1}} \right)} \right\}}}{{M{N_b}}}. \label{eq:noisevarhat_H1}
\end{align}
These MLEs are also applicable to the energy detector in \ref{sec:ED} for the case where the noise power is unknown \emph{a priori}, since the ED test threshold is a function of $\sigma _b^2$.

The log-GLRT is then obtained by substituting \eqref{eq:noisevarhat_H0}--\eqref{eq:noisevarhat_H1} into \eqref{eq:L0}--\eqref{eq:L1}:
\begin{align}
{T_{{G_1}}}\left( {{{\mathbf{Y}}_b}} \right) &= {{{\mathcal{L}_1}\left( {{{\mathbf{Y}}_b};\hat \sigma _{b|{H_1}}^2} \right)}}-{{{\mathcal{L}_0}\left( {{{\mathbf{Y}}_b};\hat \sigma _{b|{H_0}}^2} \right)}} \mathop \gtrless \limits_{{{H}_0}}^{{{H}_1}} {\eta _1}\\
& = \frac{{\Tr\left\{ {{{\left( {{{\mathbf{Y}}_b} - {{\mathbf{M}}_A}} \right)}^H}\left( {{{\mathbf{Y}}_b} - {{\mathbf{M}}_A}} \right)} \right\}}}{{\Tr\left\{ {{{\left( {{{\mathbf{Y}}_b} - {{\mathbf{M}}_1}} \right)}^H}\left( {{{\mathbf{Y}}_b} - {{\mathbf{M}}_1}} \right)} \right\}}} \mathop \gtrless \limits_{{{H}_0}}^{{{H}_1}} {e^{{{{\eta _1}} \mathord{\left/ {\vphantom {{{\eta _1}} {M{N_b}}}} \right.
 \kern-\nulldelimiterspace} {M{N_b}}}}}\label{eq:TG1_Yb}.
\end{align}
Determining the appropriate threshold $\eta \triangleq {e^{{{{\eta _1}} \mathord{\left/ {\vphantom {{{\eta _1}} {M{N_b}}}} \right.
 \kern-\nulldelimiterspace} {M{N_b}}}}}$ to meet a target $P_{FA}$ requires the pdf of the GLRT test statistic under the null hypothesis $H_0$.
Let
\begin{align}
\mathbf{X}&\triangleq \mathbf{Y}_b-\mathbf{M}_A\\
{\mathbf{W}} & \triangleq {\mathbf{X}}{{\mathbf{X}}^H}; \quad {{\mathbf{W}}_1} \triangleq \left( {{\mathbf{X}} - {{\mathbf{M}}_E}} \right){\left( {{\mathbf{X}} - {{\mathbf{M}}_E}} \right)^H}
\end{align}
based on which we can rewrite
\[{T_{{G_1}}}\left( {{{\mathbf{Y}}_b}} \right) = \frac{{\Tr\left\{ {\mathbf{W}} \right\}}}{{\Tr\left\{ {{{\mathbf{W}}_1}} \right\}}}.\]
Under the null hypothesis, we have ${\mathbf{X}} \sim \mathcal{CN}\left( {{\mathbf{0}},\sigma _b^2{\mathbf{I}}} \right)$ and thus ${\mathbf{W}}$ is a central Wishart matrix. Clearly, the matrix $\left({\mathbf{X}} - {{\mathbf{M}}_E}\right)$  is distributed as $\left( {{\mathbf{X}} - {{\mathbf{M}}_E}} \right) \sim \mathcal{CN}\left( { - {{\mathbf{M}}_E},\sigma _b^2{\mathbf{I}}} \right)$ under $H_0$, thus ${{\mathbf{W}}_1}$ in the denominator of ${T_{{G_1}}}\left( {{{\mathbf{Y}}_b}} \right)$ has a noncentral Wishart distribution.

Therefore, under hypothesis $H_0$ ${T_{{G_1}}}$ is the ratio of two dependent random variables: the trace of the central Wishart matrix $\mathbf{W}$ and trace of the noncentral Wishart matrix ${\mathbf{W}}_1$. Since this does not correspond to a known distribution and no straightforward method\footnote{The trace of a non-central Wishart matrix has a distribution characterized by zonal polynomials or infinite series.} exists to derive the exact pdf, we approximate the distribution of ${T_{{G_1}}}$ as follows.
For tractability, we first approximate $\mathbf{W}_1$ with a spatially-correlated central Wishart matrix $\mathbf{C} \mathbf{C}^H$, where ${\mathbf{C}} = {{\mathbf{\Psi }}^{{1 \mathord{\left/
 {\vphantom {1 2}} \right.
 \kern-\nulldelimiterspace} 2}}}{\mathbf{X}}$, which has approximately the same first- and second-order moments as $\mathbf{W}_1$ \cite{Tan82,Nossek10}. This yields ${\mathbf{\Psi }} = {\mathbf{I}} + {M^{ - 1}}{{\mathbf{M}}_E}{\mathbf{M}}_E^H$ as the effective correlation matrix. Since the detection of a weak LO leakage signal requires that the number of samples $M$ be many orders of magnitude larger than $N_b$, it is sufficient to consider the case of $\rank(\mathbf{Y}_b)=\rank(\mathbf{X})=N_b$. Furthermore,
 \begin{equation}\label{eq:EigIneq}
 \Tr\left\{ {{{\mathbf{W}}_1}} \right\} \simeq \Tr\left\{ {{\mathbf{\Psi X}}{{\mathbf{X}}^H}} \right\} \leq \sum\nolimits_{i = 1}^{{N_b}} {{\lambda _i}\left( {\mathbf{\Psi }} \right){\lambda _i}\left( {\mathbf{W}} \right)}
 \end{equation}
 where the eigenvalue inequality is due to \cite[Thm. 2]{Zhang_TAC}. Defining the ordered eigenvalues ${{\gamma _i} = {\lambda _i}\left( {\mathbf{W}} \right)}$ and ${{\psi _i} = {\lambda _i}\left( {\mathbf{\Psi }} \right)}$, we have
 \begin{align}
  \Pr \left\{ {{T_{{G_1}}}\left( {{{\mathbf{Y}}_b}} \right) \geq \eta } \right\} & \approx \Pr \left\{ {\frac{{\sum\nolimits_{i = 1}^{{N_b}} {{\gamma _i}} }}{{\sum\nolimits_{i = 1}^{{N_b}} {{\psi _i}{\gamma _i}} }} \geq \eta } \right\} \label{eq:step1}\hfill \\
   & \leq \Pr \left\{ {\frac{{{N_b}{\gamma _1}}}{{\sum\nolimits_{i = 1}^{{N_b}} {{\psi _i}{\gamma _i}} }} \geq \eta } \right\}\\
   & \leq \Pr \left\{ {\frac{{{N_b}{\gamma _1}}}{{\psi_{N_b}}{\sum\nolimits_{i = 1}^{{N_b}} {{\gamma _i}} }} \geq \eta } \right\} \;  \label{eq:PFAUB}
\end{align}
where \eqref{eq:step1} follows from \eqref{eq:EigIneq}.
Let ${{T_0 \triangleq {N_b\gamma _1}} \mathord{\left/
 {\vphantom {{T \triangleq {\gamma _1}} {\sum\nolimits_i {{\gamma _i}} }}} \right.
 \kern-\nulldelimiterspace} {\sum\nolimits_i {{\gamma _i}} }}$ represent the scaled largest eigenvalue of $\mathbf{W}$ divided by its trace. A number of different results are available in the literature for exact and approximate probability distributions of $T_0$. We adopt the approximate CDF in \cite[eq. 27]{Tirkonnen11} due to its accuracy and relatively simple closed-form expression, which has the form
\begin{equation}\label{eq:EigenvalueCDF}
{F_{{T_0}}}\left( y \right) = c\left( {B\left( y \right) - B\left( 1 \right)} \right),\quad y \in \left[ {1,\infty } \right]
\end{equation}
where $c = \frac{{\Gamma \left( {{m \mathord{\left/ {\vphantom {m 2}} \right.
 \kern-\nulldelimiterspace} 2}} \right){{\left( {N_b\varpi } \right)}^{ - k}}}}{{k\Gamma \left( {{m \mathord{\left/
 {\vphantom {m 2}} \right.
 \kern-\nulldelimiterspace} 2} - k} \right)\Gamma \left( k \right)}}$, $m=2MN_b$, $k$ and $\varpi$ are constants that are functions of the matrix dimensions $M$ and $N_b$ \cite[eqs. 12,13]{Tirkonnen11}, and $B\left( x \right) = {}_2{F_1}\left( {k,1 + k - 0.5m;k + 1;\frac{x}{{N_b\varpi }}} \right){x^k}$.
The desired upper bound on $P_{FA}$ in \eqref{eq:PFAUB} then simplifies to
\begin{equation}
\Pr \left\{ {\frac{{{N_b}{\gamma _1}}}{{\psi_{N_b}}{\sum\nolimits_{i = 1}^{{N_b}} {{\gamma _i}} }} \geq \eta } \right\}=1-{F_{{T_0}}}\left( \psi_{N_b}\eta \right).
\end{equation}

To compute the probability of detection for the GLRT with unknown noise variance, we must determine the distribution of $T_0$ under the alternative hypothesis $H_1$. Returning to \eqref{eq:TG1_Yb}, we now define
\[\mathbf{X}\triangleq \mathbf{Y}_b-\mathbf{M}_1,\]
 such that
 \[{T_{{G_1}}}\left( {{{\mathbf{Y}}_b}} \right) = \frac{{\Tr\left\{ {{{\mathbf{W}}_1}} \right\}}}{{\Tr\left\{ {\mathbf{W}} \right\}}}.\]
 Next, we observe ${\mathbf{X}} \sim \mathcal{CN}\left( {{\mathbf{0}},\sigma _b^2{\mathbf{I}}} \right)$, which implies that ${\mathbf{W}} = {\mathbf{X}}{{\mathbf{X}}^H}$ in the denominator of ${T_{{G_1}}}\left( {{{\mathbf{Y}}_b}} \right)$ is a central Wishart matrix. Now, the matrix ${\mathbf{X}} + {{\mathbf{M}}_E}$ is distributed as $\left( {{\mathbf{X}} + {{\mathbf{M}}_E}} \right) \sim \mathcal{CN}\left( { {{\mathbf{M}}_E},\sigma _b^2{\mathbf{I}}} \right)$, such that ${{\mathbf{W}}_1} = \left( {{\mathbf{X}} + {{\mathbf{M}}_E}} \right){\left( {{\mathbf{X}} + {{\mathbf{M}}_E}} \right)^H}$ in the numerator of ${T_{{G_1}}}\left( {{{\mathbf{Y}}_b}} \right)$ is a non-central Wishart matrix. Therefore, under hypothesis $H_1$, ${T_{{G_1}}}$ is the ratio of two dependent random variables: the trace of the noncentral Wishart matrix $\mathbf{W}_1$ and trace of the central Wishart matrix ${\mathbf{W}}$. Similar to \eqref{eq:EigIneq} we set $\Tr\left\{ {{{\mathbf{W}}}} \right\} \simeq \Tr\left\{ {{\mathbf{\Psi X}}{{\mathbf{X}}^H}} \right\}$, where $\mathbf{\Psi}={\mathbf{I}} + {M^{ - 1}}{{\mathbf{M}}_E}{\mathbf{M}}_E^H$ is the effective correlation matrix. We can then repeat the preceding steps involved in the computation of $P_{FA}$ with minor modifications, to obtain the following lower bound on the detection probability:
\begin{equation}
\Pr \left\{ {{T_{{G_1}}}\left( {{{\mathbf{Y}}_b}} \right) \geq \eta } \mid H_1\right\}\geq {F_{{T_0}}}\left( \frac{\varsigma_{N_b}}{\eta} \right) \; ,
\end{equation}
where $\varsigma_{N_b}$ is the smallest eigenvalue of $\mathbf{\Psi}$, and $F_{{T_0}}\left( \cdot \right)$ is defined in \eqref{eq:EigenvalueCDF}.

\subsection{Unknown noise variance and leakage channel of Eve}\label{sec:GLRT2}
We now consider the most general case where the unknown parameters are the noise variance and Eve's leakage amplitude and phase, i.e., $\left( {\sigma _b^2,{{\mathbf{H}}_{be}},\{A_{{i}}\}_{i=1}^{N_e},\{\theta_i[n]\}_{i,n} }\right)$. Recall that Alice and Bob are cooperative nodes and can exchange estimates of Alice's leakage parameters, which we continue to assume are known at Bob.
Since frequency $\omega$ is known, we can rewrite \eqref{eq:M_E} as ${{\mathbf{M}}_E} = {{{\mathbf{M'}}}_E}{\mathbf{D}}$, where ${{{\mathbf{M'}}}_E}$ is unknown and ${\mathbf{D}} = \diag\left\{ {1,{e^{j{\omega}}}, \ldots ,{e^{j{\omega}\left( {M - 1} \right)}}} \right\}$ is a known full-rank matrix.
Direct differentiation of \eqref{eq:TG1_Yb} yields the MLE (cf. \cite{Scharf05})
\begin{equation}\label{eq:M_hat}
{{{\mathbf{\hat M'}}}_E} = \left( {{{\mathbf{Y}}_b} - {{\mathbf{M}}_A}} \right){{\mathbf{D}}^H}.
\end{equation}

The log-GLRT is then given by ${T_{{G_2}}}\left( {{{\mathbf{Y}}_b}} \right) \mathop \gtrless \limits_{{{H}_0}}^{{{H}_1}} {e^{{{{\eta _2}} \mathord{\left/
 {\vphantom {{{\eta _2}} {M{N_b}}}} \right.
 \kern-\nulldelimiterspace} {M{N_b}}}}}$, where
\begin{equation}\label{eq:TG2}
{T_{{G_2}}}\left( {{{\mathbf{Y}}_b}} \right) = \frac{{\Tr\left\{ {{{\left( {{{\mathbf{Y}}_b} - {{\mathbf{M}}_A}} \right)}^H}\left( {{{\mathbf{Y}}_b} - {{\mathbf{M}}_A}} \right)} \right\}}}{{\Tr\left\{ {{{\left( {{{\mathbf{Y}}_b} - {{{\mathbf{\hat M'}}}_E}{\mathbf{D}} - {{\mathbf{M}}_A}} \right)}^H}\left( {{{\mathbf{Y}}_b} - {{{\mathbf{\hat M'}}}_E}{\mathbf{D}} - {{\mathbf{M}}_A}} \right)} \right\}}}.
\end{equation}
Substituting \eqref{eq:M_hat} into \eqref{eq:TG2}, we have
\begin{equation}\label{eq:TG2step2}
{T_{{G_2}}}\left( {{{\mathbf{Y}}_b}} \right) = \frac{{\Tr\left\{ {{{\left( {{{\mathbf{Y}}_b} - {{\mathbf{M}}_A}} \right)}^H}\left( {{{\mathbf{Y}}_b} - {{\mathbf{M}}_A}} \right)} \right\}}}{{\Tr\left\{ {{\mathbf{M}}_A}^H{{\mathbf{M}}_A} \right\}}}.
\end{equation}
Therefore, we can utilize existing expressions for the distribution of the trace of a central and non-central Wishart matrix to compute the exact $P_{FA}$ and $P_D$ of the above test, respectively. The trace of a central Wishart matrix follows a (scaled) chi-squared distribution, thus
\begin{equation}
{P_{FA}} = 1-\mathcal{P}\left( {M{N_b},\sigma _b^{ - 2}\Tr\left\{ {{\mathbf{M}}_A^H{{\mathbf{M}}_A}} \right\}{e^{{{{\eta _2}} \mathord{\left/
 {\vphantom {{{\eta _2}} {M{N_b}}}} \right.
 \kern-\nulldelimiterspace} {M{N_b}}}}}} \right)
\end{equation}
where $\mathcal{P}\left(\cdot,\cdot\right)$ is the regularized Gamma function. Under $H_1$, \eqref{eq:TG2step2} is the scaled trace of a non-central Wishart matrix (equivalent to a weighted sum of independent non-central chi-squared variables), the exact distribution of which is given in \cite{Kourouklis85}. Thus,
\begin{equation}
P_D= 1-\sum\limits_{k = 0}^\infty  {{c_k}} \mathcal{P}\left( {M{N_b} + k,\frac{1}{2\lambda\sigma _b^{2}}\Tr\left\{ {{\mathbf{M}}_A^H{{\mathbf{M}}_A}} \right\}{e^{{{{\eta _2}} \mathord{\left/
 {\vphantom {{{\eta _2}} {M{N_b}}}} \right.
 \kern-\nulldelimiterspace} {M{N_b}}}}} } \right)
\end{equation}
where coefficients $c_k$ are computed recursively based on \cite[eqs. 2.6,2.9]{Kourouklis85}, and $0<\lambda<\infty$ is arbitrary. A simpler expression can be obtained if desired by approximating the above trace with a single non-central chi-squared variate of the same first moment \cite{Barton89}.

\section{Eavesdropping Strategies}\label{sec:EveStrats}
We have thus far proposed and characterized various statistical tests to determine the presence of a passive MIMO eavesdropper. In the event that the eavesdropper is a malevolent adversary, she may take evasive measures to avoid detection while at the same time attempting to intercept as much information as possible. This implies that Eve is omniscient in the sense that she is able to estimate the relative locations of Alice/Bob and the leakage detection scheme in place. The key parameters under Eve's control are her location relative to Alice/Bob and the number of antennas $N_e$: decreasing $d_{ae}$ or increasing $N_e$ enhances both her interception rate as well as the likelihood of being detected.

First consider the impact of Eve's distance from Alice and Bob for a given $N_e$. Let us define the instantaneous leakage rate from Alice to Eve as
\begin{equation}\label{eq:R_e}
R_e ={\log _2}\left| {{\mathbf{I}} + {{\mathbf{H}}_{ea}}{\mathbf{QH}}_{ea}^H}{\mathbf{Z}}_e^{ - 1} \right|,
 \end{equation}
which is the second term of the MIMO secrecy rate expression in \eqref{eq:MIMOSecRate}. It is assumed that Eve is interested in maintaining a threshold $\bar{R}_e$ for the instantaneous or average leakage rate. For Rayleigh fading channels, a Gaussian approximation for the MIMO mutual information in the large-antenna regime \cite[Thm. 3]{Tarokh04} results in the following average leakage rate:
\begin{equation*}
\mathcal{E}_{\mathbf{H}_{ea}}\left\{ {{R_e}} \right\} \simeq {d_{ae}^{-\alpha}}{N_e}\frac{{{P_a}}}{{{N_a}}}{\log _2}e.
\end{equation*}

As an example, the eavesdropper can seek to minimize the probability of being detected, subject to a minimum average leakage rate constraint:
\begin{equation}\label{eq:minPdc}
\begin{split}
(\textrm{P1}):\quad \min_{d_{ae},d_{be}} &\quad P_{dc}  \\
\mathrm{s.t.} & \quad\mathcal{E}\{R_e\} \geq {{\bar R}_e}
%& \quad \mathbf{Q}_s \succ \mathbf{0}.
\end{split}
\end{equation}
This problem has a straightforward solution.
 An average leakage rate constraint of ${\bar R}_e$ requires
\[{{d}_{ae}} \leq \left(\frac{{{{\bar R}_e}{N_a}{P_a}}}{{{N_e}{{\log }_2}e}}\right)^{-1/\alpha} \; . \]
Since the leakage rate constraint is independent of $d_{be}$, and since $P_{dc}$ is decreasing with respect to $d_{ae}$, the optimal value of $d_{ae}$ is attained when the leakage rate constraint is met with equality. The optimal value of $d_{be}$ is then the furthest possible distance from Bob for this optimal $d_{ae}^*$. Geometrically, Eve's optimal location is the point of intersection of the circle of radius $d_{ae}^*$ centered at Alice and the line joining Alice and Bob.

The LO leakage phenomenon essentially precludes the use of fast antenna switching at Eve, therefore she must decide on how many antennas are to be deployed at the initiation of the eavesdropping attack. A rough assessment of her risk-reward tradeoff can be obtained by simply enumerating $\bar R_s$ for the various possible values of $N_e$. Further insight can be obtained by analyzing the behavior of $P_{dc}$, $R_e$, and $\bar R_s$ as $N_e$ varies: does the rate at which $P_{dc}$ increases outweigh the rate of gain in $R_e$ or vice versa? This exercise is carried out via simulations in Sec.~\ref{sec:sim}.

\section{Numerical Results}\label{sec:sim}
We present simulation results obtained by averaging over 1000 i.i.d. Rayleigh channel fading instances for several network scenarios. In each instance the eavesdropper is present with probability $\beta=0.5$. Unless stated otherwise, we set the number of antennas as $N_a=N_b=N_e=4$, the distance between Alice and Bob is assumed to be $d_{ab}=10$m, and $d_{ae}=d_{be}$. The leakage amplitude is set to -50 dBm/antenna with an IF frequency of 200 kHZ and unit noise power for all users, and the number of samples is fixed at $M=10^5$.

\begin{figure}[htbp]
\centering
\includegraphics[width=0.96\linewidth]{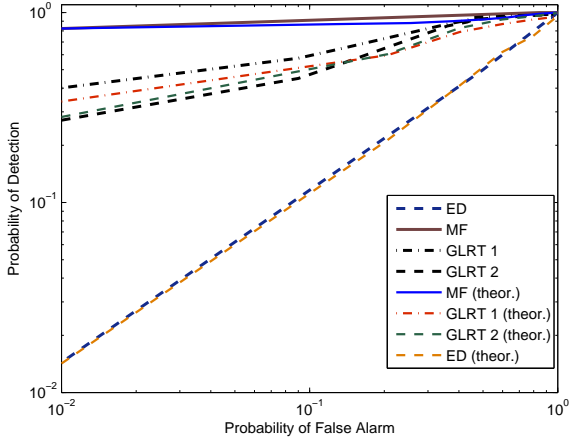}
\caption{ROC performance of various eavesdropper detectors.}\label{fig:ROC}
\end{figure}

The empirical and theoretical receiver operating characteristic (ROC) curves of the detection schemes described in Sec.~\ref{sec:ED_MF} and Sec.~\ref{sec:GLRTs} are displayed in Fig.~\ref{fig:ROC}. GLRT 1 and GLRT 2 denote the composite tests described in Sec.~\ref{sec:GLRT1} and Sec.~\ref{sec:GLRT2}, respectively. The detection probabilities shown here are the local metrics at Bob, and the theoretical results are generally in good agreement with simulations. The energy detector fails to distinguish between the null and alternative hypotheses and is virtually unusable, even under the assumption of a perfectly known noise variance $\sigma_b^2$. The GLRTs are both more robust and perform much closer to optimal MF detection as compared to ED. The joint detection probability $P_{dc}$ in \eqref{eq:Decision_fusion} will by definition exhibit similar trends.

\begin{figure}[htbp]
\centering
\includegraphics[width=\linewidth]{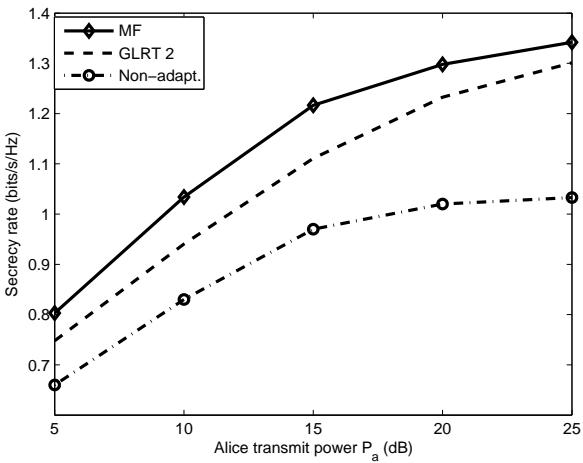}
\caption{Ergodic secrecy rate versus transmit power $P_a$.} \label{fig:Rs}
\end{figure}
Fig.~\ref{fig:Rs} depicts ${{\bar R}_s}$ versus Alice's total power constraint $P_a$ for ED and GLRT 2 detectors, as well as a non-adaptive scheme which pessimistically assumes that Eve is always present and always has Alice allocate a fraction of power for artificial noise \cite{GoelN08}. Eve is located 10m away from both legitimate terminals. The local eavesdropper detection decisions at Alice and Bob are combined using an OR fusion rule. It is evident that the eavesdropper detection schemes outperform the non-adaptive strategy by reducing the unnecessary allocation of resources for secure transmission when the eavesdropper is absent.

\begin{figure}[htbp]
\centering
\includegraphics[width=0.9\linewidth]{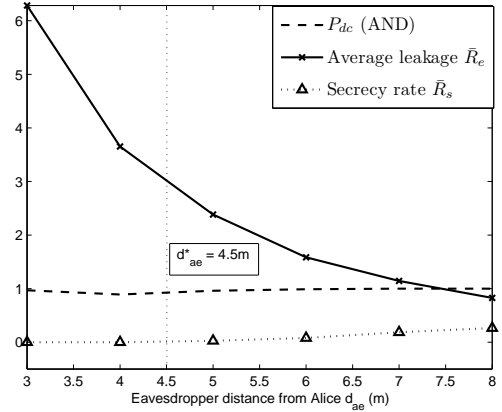}
\caption{Rates and joint detection probability as a function of distance $d_{ae}$.}\label{fig:dist}
\end{figure}

In Fig. \ref{fig:dist}, the eavesdropper is moved along a line parallel to the line between Alice and Bob, with $d_{ab}=9$m being fixed. It is assumed Eve has a desired leakage rate target of $\bar R_e =3$, which translates into an optimal distance of roughly $d_{ae}=4.5m$ according to \eqref{eq:minPdc}. This predicted distance is quite close to the observed value of $d_{ae}$ corresponding to the empirical rate $\bar R_e = 3$. Furthermore, the  joint detection probability using MF is at its lowest value around this spatial location, which is intuitive since increasing $d_{ae}$ brings Eve closer to Bob, thus $\bar R_s$ is seen to increase with $d_{ae}$ due to the combined factors of improved $P_{dc}$ and diminishing leakage $R_e$.
From the perspective of the legitimate nodes, increasing the number of observation samples $M$ to further improve $P_{dc}$ detracts from the time available for data transmission, while increasing $N_a$ or $N_b$ will improve the interception capability of Eve. On the other hand the interception capability of Eve is degraded as $d_{ae}$ grows; the interplay of these factors has interesting implications for the eavesdropper when she can choose where to position herself.

\begin{figure}[htbp]
\centering
\includegraphics[width=0.94\linewidth]{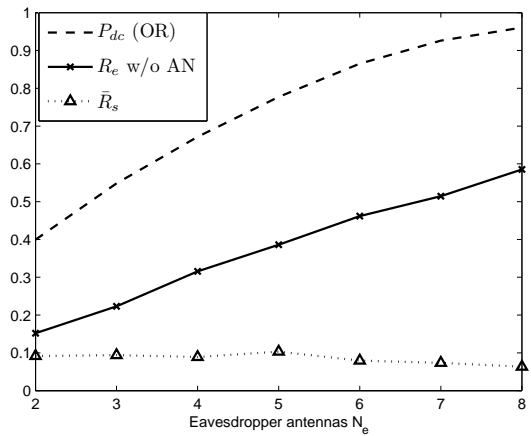}
\caption{Joint detection probability and leakage/secrecy rates versus number of eavesdropper antennas.}\label{fig:EveAnt}
\end{figure}

Fig.~\ref{fig:EveAnt} highlights the joint detection probability based on MF, along with leakage rate $R_e$ and average secrecy rate $\bar R_s$, as $N_e$ varies for fixed $N_a=N_b=2$. In a conventional MIMO system the received mutual information is always non-decreasing in the number of receiver antennas, and this is borne out by the behavior of $R_e$. On the other hand, in our wiretap setting an increase in $N_e$ also boosts the ability of the legitimate nodes to detect the eavesdropper, as exemplified by $P_{dc}$. As a result, the average secrecy rate actually increases as $N_e$ grows from two to five, while ultimately approaching zero in the regime where Eve has four times as many antennas as Alice and Bob. Thus, an interesting tradeoff exists for Eve when choosing how many antennas to activate for wiretapping.

\section{Conclusions}\label{sec:Conclu}
In the MIMO wiretap channel, it is critical that the presence of a passive eavesdropper be determined so as to enable robust secrecy-encoding schemes as a countermeasure. In this work we studied the performance of methods in which the legitimate nodes attempt to detect the eavesdropper from the local oscillator power that is inadvertently leaked from its RF front end. We analyzed the performance of non-coherent energy detection as well as optimal coherent detection to obtain lower and upper limits on the achievable detection probability. Subsequently, two robust detectors based on GLRTs were derived to account for cases with unknown leakage and noise parameters. We then showed how the proposed detectors allow the legitimate nodes to increase the MIMO secrecy rate of the channel. Issues of interest for further study include the design of sequential detectors for optimizing the sensing duration used to detect potential eavesdroppers.

\end{document}